\documentstyle[epsf,twocolumn,prl,aps]{revtex}
\begin{document}

\twocolumn[\hsize\textwidth\columnwidth\hsize\csname
@twocolumnfalse\endcsname

\title{On the Nagaoka polaron in the $t-J$ model}
\author{Steven R. White$^{1}$ and Ian Affleck$^{2}$}
\address{$^1$Department of Physics and Astronomy, University of California,
         Irvine, CA 92697 USA\\}
\address{$^2$Canadian Institute for Advanced
Research  and Department of Physics and Astronomy, University of 
British Columbia, 
Vancouver, B.C., Canada, V6T 1Z1 \\ }
\maketitle
\begin{abstract}
It is widely believed that a single hole in the two (or three) dimensional
$t-J$ model, for sufficiently small exchange coupling $J$, creates a
ferromagnetic bubble around itself, a finite $J$ remnant of the
ferromagnetic groundstate at $J=0$ (the infinite $U$ Hubbard model), first
established by Nagaoka \cite{nagaoka}.
We investigate this phenomenon in two dimensions using the density matrix
renormalization group, for system sizes up to $9\times 9$.  We find 
that the polaron forms 
for $J/t<0.02-0.03$ (a somewhat larger value than estimated previously).  
Although finite-size effects appear large, our data seems consistent 
with the expected $1.1(J/t)^{-1/4}$ variation of polarion radius.  We also 
test the Brinkman-Rice model of non-retracing paths in a N\'eel background, 
showing that it is quite accurate, at larger $J$.  Results are 
also presented in the case where the
 Heisenberg interaction is dropped (the $t-J^z$ model).
Finally we discuss a ``dressed polaron'' picture in which the hole 
propagates freely inside a finite region but makes only self-retracing
excursions outside this region.
\end{abstract}

% comment out the following line for single column output
 ]

An understanding of the groundstate of the two dimensional  $t-J$ model near
half-filling remains
one of the most urgent open problems in condensed matter theory. 
The Hamiltonian may be written 
\begin{equation}
H=\sum_{<i,j>}\left[ -tP(\sum_{\alpha}c^\dagger_{i\alpha}c_{j\alpha}+h.c.)P
+J(\vec S_i\cdot \vec S_j-n_in_j/4)\right].\end{equation}
Here $h.c.$ stands for Hermitean conjugate, $c_{i\alpha}$ annihilates
an electron of spin $\alpha$ at site $i$, $P$ projects out states with
no double occupancy on any site, $<i,j>$ represents pairs of
nearest neighbor sites, and:
\begin{equation}
\vec S_i\equiv \sum_{\alpha,\beta}
c^\dagger_{i\alpha}{\vec \sigma_{\alpha \beta} \over 2}c_{i\beta}.
\end{equation}
Even the 
case of a single hole relative to half-filling is highly non-trivial.  At $J=0$, corresponding
to the infinite $U$ Hubbard model, the groundstate was proved to be
ferromagnetic by Nagaoka.\cite{nagaoka}
  This happens because the hole can move around
most efficiently, thus minimizing the kinetic energy, when the spins
are fully polarized.  Conversely, an antiferromagnetic exchange coupling, $J$, favours
anti-parallel spins leading to the picture that a bubble of polarized 
spins forms around the hole while the spins further away are in the antiferromagnetic
groundstate.  The hole essentially moves around freely only inside this bubble,
or polaron, longer range motion leading to disruption of the antiferromagnetic
groundstate.  However, this picture is only expected to hold, at best, for
sufficiently small $J/t$.  At somewhat larger $J/t$ quite different
behavior is expected, perhaps the Brinkman-Rice picture in which the hole
makes only self-retracing excursions from the origin in order to avoid 
disrupting the antiferromagnetic background.  
At values of $J$ relevant to the cuprates, $J/t \sim 0.3-0.4$,
numerical studies have demonstrated the presence of frustrating
singlet correlations across mobile holes, connecting sites on
the same sublattice\cite{holestructures}. 
However, a simple analytic approach reproducing this
effect is lacking.

Recent advances in numerical
methods now make it possible to study the one hole problem accurately.  
We present here
density matrix renormalization group\cite{dmrg} (DMRG) results for the small $J/t$
regime on lattices of size 
up to $9\times 9$
with boundary conditions that are either open or mixed 
open and periodic (cylindrical).  
The number of states kept per block ranged from 2000 to 4000.
Various methods of studying the Nagaoka polaron are possible.  We
 study the nearest neighbour exchange energy 
$\langle\vec S_{\vec r}\cdot \vec S_{\vec r+\hat e}-n_{\vec r}n_{\vec r+\hat e}/4\rangle$,
 as a function of position $\vec r$.
This is increased in the vicinity of the hole, which resides near the center of
the lattice due to a combination of boundary conditions and initial conditions
in the DMRG calculation.  We calculate the groundstate energy as a function of
$J/t$, showing reasonable agreement with the $\sqrt{J}$ dependence expected
in the polaronic state, reviewed below.  Most directly, we can calculate the
total spin of the groundstate with an even number of sites and a single hole.  
We find that this varies from $1/2$ for $J/t>0.02-0.03$, where the polaron disappears, 
to $(N-1)/2$, where $N$ is the number of sites for $J/t \sim .001$.
In this latter case the polaron becomes larger
than the lattice size.  

The energy and radius of the Nagaoka polaron, as a function of $J/t$
are derived by assuming 
that the hole wave-function is that of a free particle with
vanishing boundary conditions at the edge of the circular bubble.  The
groundstate is given by the lowest energy such state.  R is
determined by balancing the  kinetic energy of the hole
against the energy cost of the ferromagnetic bonds.  For small J
and hence large R the hole wave-function corresponds to a small
wave-vector, at the bottom of the band, which can be approximated
as quadratic, with an effective mass of 1/2t.  (Henceforth we set
t=1.)  The wave-function is hence a Bessel function,
$J_0(kr)$ with k chosen so that $J_0(kR)=0$.  The first zero
of $J_0(x)$ occurs at $x\approx 2.40$.  Hence the kinetic energy, compared
to that at half-filling is:
\begin{equation}E_K=-4 + (2.40)^2/R^2.\label{E_K}\end{equation}
To this we must add the energy cost of the ferromagnetic bonds.  A ferromagnetic bond costs energy $J/2$ relative to the energy of
a bond in an up-down state.  However, the actual energy per bond of the N\'eel
groundstate for the two dimensional square lattice is approximately $-0.33J$, 
so the energy cost of a ferromagnetic bond is about $0.58J$.  
Taking into account that there are 2 bonds per unit area (we set the
lattice constant to 1) we see that the magnetic energy, measured relative
to that of the groundstate with no hole, is:
\begin{equation}
E_M = 1.16J\pi R^2.\label{E_M}\end{equation}
Thus the total energy of the polaron, relative
to the antiferromagnetic groundstate with no hole present is:
\begin{equation}
E=-4+ (2.40)^2/R^2+1.16J\pi R^2.\label{E}\end{equation}
Minimizing the energy with respect to R gives:
\begin{eqnarray}R&=&1.12 J^{-1/4}\nonumber \\
E&=&-4+9.20 \sqrt{J}.\label{E,R}\end{eqnarray}  
The total spin of the polaron is:
\begin{equation}
S=(1/2)\pi R^2=1.97J^{-1/2}\label{S(R)}.
\end{equation}

We note that, even if we assume that these formulas become exact in the limit
 $J\to 0$ for an infinite area system, we should expect severe finite size effects.
$R$ must be much bigger than 1 lattice spacing in order for the assumption of
a circular polaron with a simple vanishing boundary condition to be reasonable.  On 
the other hand, the system length must be much larger than $R$ in order for
the polaron not to be perturbed by the finite system size.  Thus, in practice 
there will be at best a narrow range of small $J$'s for which these formulas
apply in a numerical simulation.  If $J$ is too small, the polaron simply
fills the entire system.  In this case its energy is trivial to calculate.  The
 kinetic energy is that of a single hole in a filled (spin polarized) band.  If
the system has dimensions $L\times M$ with free boundary conditions,
 for example, 
this is:
\begin{equation}
E_K=-2\cos [\pi /(L+1)]-2\cos [\pi /(M+1)].\label{EKf}
\end{equation}
$\langle\vec S_i\cdot \vec S_j-n_in_j/4\rangle$ is exactly zero in this state.
The magnetic energy, relative to the no-hole ground state,
is simply $0.58$ times the number of bonds 
\begin{equation} E_M=0.58J[2LM-(L+M)],\label{EMf}\end{equation}
up to finite size corrections to the groundstate energy of the hole-free 
system.
   
We expect the polaron to fill the entire system when $J$ is reduced to 
a value of order $1/N^2$ where $N$ is the number of sites.  On the other
hand, as $J$ is increased we expect the polaron to begin showing
departures from circular shape as $R$ becomes not so much larger than 1.  
Eventually when $R$ is of O(1) the magnetization of the polaron is reduced
to $1/2$ at which point it no longer exists.  ($S=1/2$ is the smallest
possible value for one hole relative to half-filling in a system with
an even number of sites.)  Thus, as we decrease $J$ for a given large system
size we expect $S$ to increase approximately as $1.97J^{-1/2}$ eventually
saturating at $(N-1)/2$ while the energy decreases as
$-4+9.2J^{1/2}$ until it eventually crosses over to the linear $J$-dependence
of Eqs. (\ref{EKf}) and (\ref{EMf}).  Another way that the 
polaron can be studied using DMRG takes advantage of the iterative
nature of this numerical method.  We begin our DMRG sweeps with the 
hole sitting at the origin.  In principle, after an infinite number
of DMRG sweeps in a periodic system, it should go into an extended 
state where it spreads over the entire lattice.  In this case
the groundstate should be a momentum eigenstate
in which observables like the magnetic energy density are translationally
invariant.  However, in practice, we use open or cylindrical boundary
conditions. For open BCs, the boundaries tend to localize the
polaron in the center of the system. Even with cylindrical BCs, 
we find that the polaron tends to
remain localized at the origin.  The fact that this occurs, while the
expectation value of the Hamiltonian in the approximate groundstate
appears to be close to its minimum, indicates that the polaron has a 
large effective mass.  Since the polaron remains well localized we
can study its size by measuring spin correlations,
$\langle\vec S_{\vec r}\cdot \vec S_{\vec r+\hat e}-n_{\vec r}n_{\vec r+\hat e}\rangle$,
as a function of $\vec r$.

These measures of the polaron size versus $J$ are shown in the figures.
In Fig. 1 we show the kinetic and magnetic energy versus $J$.  Note
that the agreement with Eqs. (\ref{E_K}) and (\ref{E_M}) is fair for
$J<0.02-0.03$ but that the data starts to deviate strongly from the
theoretical curve 
at this point, suggesting the breakdown of the polaron picture.  
The sharp features in $E_M$ and $E_K$ at $J \approx 0.03$ are
rather surprising and may be numerical artifacts.  The
uncertainties in measurements of the magnetic and kinetic energies
separately are much larger than those associated with the total
energy. We have
included rough error bars, estimated using the energy measurements as
a function of the number of states kept per block, and 
have also shown some data for a $9
\times 9$ lattice for this reason. The $9 \times 9$ results do
not clearly show a local minimum in $E_M$ near 0.03.
Regardless of whether there is such nonmonotonic behavior in
$E_M$ or $E_K$,
there seems to be rather sharp behavior associated with an
abrupt crossover as the polaron disappears.
We have
also measured the spin of the polaron for various $J$, shown in Fig. 2.
While $S$ certainly increases with decreasing $J$, eventually reaching
saturation, at $J\approx 0.001$, it seems to lie well below the polaron
prediction of Eq. (\ref{S(R)}), at all $J$.

Finally, in Fig. 3 we present spin correlations, showing a disturbance of
the antiferromagnetic state in the vicinity of the hole, situated at
the origin.   The region over which a significant departure from 
the antiferromagnetic state occurs gives another measure of the size of
the polaron. The polaron does indeed look approximately circular, as expected. 

Taken together, we think that this data suggests rather strongly that the
Nagaoka polaron does indeed form, and that Eqs. (\ref{E_K}) and (\ref{E_M})
are asymptotically correct, for small enough $J$ and an infinite lattice.  
We have also shown that the value of $J$ where this polaron picture
breaks down is about $0.02-0.03$, five times larger than a previous analytical
estimate \cite{SS}.  However, this value is still about ten times smaller than
that believed to be relevant to the cuprates.

\begin{figure}[ht]
\epsfxsize=2.5 in\centerline{\epsffile{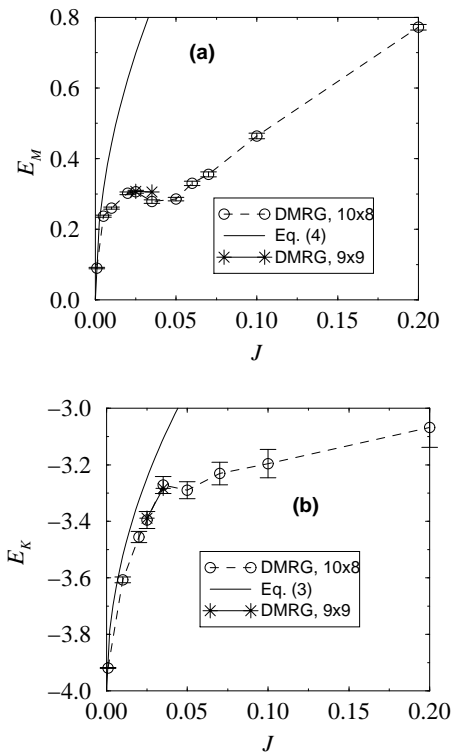}}
\caption{
(a) The magnetic energy $E_M$ of a single hole as a function of
$J$.
The continuous line is Eq. (4), using Eq. (6).
(b) The kinetic energy $E_K$. Here, the continuous line is Eq.  (3),
using Eq. (6).
}
\end{figure}         

\begin{figure}[ht]
\epsfxsize=2.5 in\centerline{\epsffile{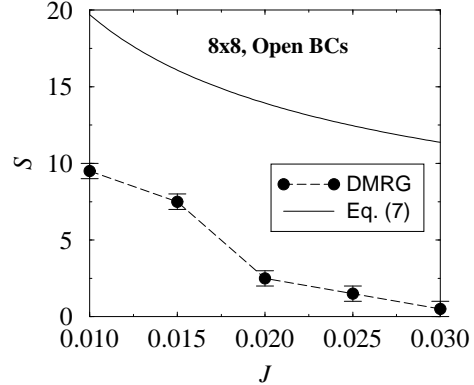}}
\caption{
The total spin of the ground state of an $8\times8$ system with
one hole, as a function of $J$. The spin was determined by
looking for degeneracy in the ground state energy for different
DMRG calculations as we very $S_z$. The result for $S$ is the
largest value of $S_z$ degenerate with $S_z=1/2$.
}
\end{figure}           
\begin{figure}[ht]
\epsfxsize=2.5 in\centerline{\epsffile{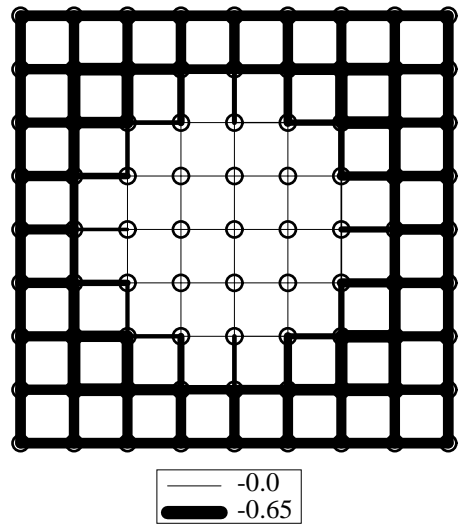}}
\caption{
The expectation value of $\vec S_i\cdot \vec S_j-n_in_j/4$,
displayed as the thickness of each line, for a $9\times9$ open
system with one hole and $J=0.01$. The ferromagnetic polaron is clearly
visible in the central region, where the expecation value is
nearly zero.
}
\end{figure}         
We have also compared the energy to that of the Brinkman-Rice 
approximation,\cite{BR} 
in the form used by Shraiman and Siggia\cite{SS} (BRSS approximation). 
This approximation should
be better for the $t-J^z$ model, in which the spin-flip part of the
Heisenberg interaction is dropped.  Then as the hole propagates it
leaves behind a string of flipped spins.  
If the hole only propagates along the $\hat x$ axis, for example, then
after it has moved $r$-steps there are $(2r+1)$ ferromagnetic bonds
 so the exchange energy is simply: 
\begin{eqnarray}
V(r)&=&(5/2+r)J, \ \  (r\geq 1)\nonumber \\
V(0) &=& 2J.\label{pot}
\end{eqnarray}
 (Note that we 
include the $-n_in_j/4$ term in the exchange energy.)  In the BRSS
approximation, the potential energy is assumed to have this value
for all strings of length $r$, regardless of their shape. 
However, if the 
hole moves around a single plaquette,  then the exchange energy
is only $5J$ after three steps, rather than $(5/2+3)J$.  Thus the 
potential energy is not the same for all strings of length $3$ and larger.
Another aspect of the BRSS approximation is to treat each string as 
a distinct state.  In this case the problem becomes equivalent to 
a single partice on a Bethe lattice with a hopping term $-1$ and
the potential energy of Eq. (\ref{pot}), for all sites at depth $r$.
The hole can hop in any of 4 directions at the first step.  Thereafter, 
there are only three non-retracing directions available.  Therefore, 
the Bethe lattice has 1 state at depth $0$, 4 states at depth $1$ 
and $4\times 3^{r-1}$ states at depth $r$ for $r\geq 2$.
 However, the Bethe lattice approximation is not strictly correct, since
 there are several different paths which lead to the 
identical hole position and spin orientation.  For instance, the
same configuration is obtained when the hole goes around a single
plaquette 1-1/2 times either in a clockwise or counter-clockwise sense.
Thus the Bethe lattice doesn't have the correct connectivity.  The 
actual $t-J^z$ model would correspond to a Bethe lattice with some of
the sites identified, starting at depth 6, and potential energies 
which are not the same for all sites at a given depth.  

It is 
generally assumed, in the BRSS approximation, that the hole wave-function
has the same value at all sites of a given depth.  Then the Bethe 
lattice model becomes equivalent to a one-dimensional model with
hopping amplitude $-2$ between site 0 and 1 and 
$-\sqrt{3}$ between sites $r$ and $r+1$ for $r\geq 1$.
 The potential energy is $2J$ on site $0$ and $(5/2+r)J$ on site $r$ 
for $r\geq 1$. A free boundary condition occurs at $r=0$.  The single particle 
groundstate for this one dimensional problem can easily be found
numerically.  In particular, when $J\ll t$,  we can use a continuum 
approximation, approximating the dispersion relation as quadratic 
near the bottom of the band, giving:
\begin{equation}
H=-2\sqrt{3}-\sqrt{3}{d^2\over dx^2}+Jx,  \ \ (x>0)
\end{equation}
with a vanishing boundary condition at $x=0$.  An exact scaling
argument then implies that the groundstate energy is proportional 
to $J^{2/3}$ and the numerical solution of this eigenvalue
problem (in terms of the Airy function)  gives the energy:
\begin{equation}
E\approx -2\sqrt{3} +2.81J^{2/3}.\label{ESS}\end{equation}
For larger $J$ the continuum approximation breaks down but the
model can be solved numerically on the lattice.  Shraiman and 
Siggia\cite{SS} state 
that the numerical solution of this equation is well approximated by 
Eq. (\ref{ESS}) with the factor in the second term replaced by $2.74$ over
the range $0.005<J/t<1$. (We have verified this result.) 
In Fig. 4 we compare the DMRG results for
the groundstate energy to the BRSS approximation.  The agreement looks 
quite good  for $J>0.02-0.03$. We note that the
BRSS approximation does not give a lower bound on the energy and appears 
to lie below the actual energy as determined by DMRG.  

\begin{figure}[ht]
\epsfxsize=3.0 in\centerline{\epsffile{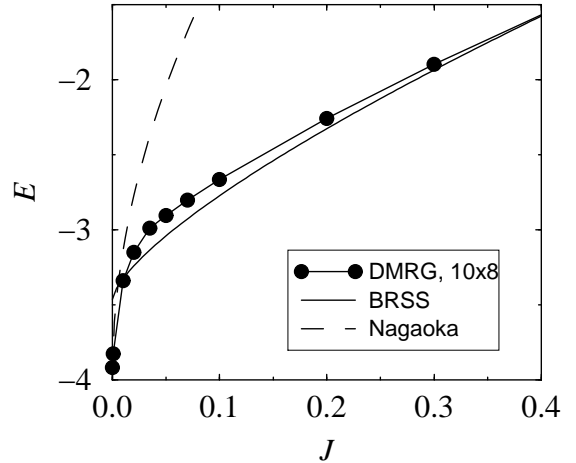}}
\caption{
The total energy of a hole, measured relative to the undoped
system,
as a function of $J$, on a $10\times8$ system.
The curve labeled BRSS is Eq. (12) with 2.81 replaced by 2.74,
and with an extra term $J$ added on to account for the term 
$-J n_in_j/4$ in the Hamiltonian which was not included in the
Hamiltonian studied by BRSS. The curve labeled Nagaoka is Eq.
(6), also with an extra $J$ added on.
}
\end{figure}

Some insight into the corrections to the polaron picture for slightly 
larger $J$ can be obtained by combining it with the BRSS string 
approximation.  We could thus consider a state where the hole moves 
in a ferromagnetic background inside a bubble of radius $R$ but also
makes excursions outside this bubble which take the form of self-retracing 
walks or strings.  Following BRSS we may approximate the lattice outside 
the polaron as a Bethe lattice.  In other words, as shown in Fig. 5,
we effectively attach 
``Bethe hair'' to each point on the boundary of the polaron so that 
the hole propagates on a square lattice inside the polaron but on 
a Bethe lattice outside the polaron.  For large $R$ (small $J$) the 
energy is close to $-4$.  This implies than when the hole enters the 
Bethe lattice it propagates below the band-edge at $-2\sqrt{3}$.  Again 
making the reasonable approximation that the wave-function has the 
same value at all points at the same depth on the Bethe lattice, the 
problem is again approximated as a one-dimensional tight-binding model 
with a linear potential.  The linear potential plays little role at
small $J$ 
since the sub-band wave-function decays exponentially even at $J=0$.
To see this note that the Schrodinger equation, everywhere on the Bethe
lattice except near the top has the form:
\begin{equation}
E\phi_r=-\sqrt{3}(\phi_{r}+\phi_{r+1}),\end{equation}
with $E\approx -4$.  The two solutions of this equation asymptotically 
behave as $\phi_r \approx 3^{\pm r/2}$.  A valid solution of the 
eigenvalue problem  must 
give only the exponentially decreasing solution, $\propto 3^{-r/2}$.
Actually solving the Schroedinger equation would require making 
small $R$-dependent adjustments in the energy (near $-4$) and
adjusting the wave-function near the edge of the polaron to ensure 
that only the exponentially decreasing solution is obtained on the 
Bethe lattice.  We might expect the Bethe lattice approximation to 
actually work better in this regime (smaller $J$ when the energy 
approaches $-4$) than at larger $J$ where the BRSS approximaton 
is normally applied ($E>-2\sqrt{3}$), because the hole has only a
 small amplitude to 
descend to large depths in the Bethe lattice, due to the exponential 
decay of the wave-function.  As noted above, the 
Bethe lattice approximation only starts to break down at fairly 
large depth (depth 6 for the lattice connectivity).  We note that
the Bethe hair may be regarded as a sort of boundary layer attached 
to the polaron.  By the above argument it only has a width of a 
few lattice spacings, independent of $R$, at large $R$.  Thus the 
naive estimates of $R(J)$ and $E(J)$ in Eq. (\ref{E,R}) should remain
true at small enough $J$.  The Bethe hair just represents a sort 
of $1/R$ correction to the polaron approximation.  
However, we see 
that it will make the finite size difficulties with numerical
simulations even more severe.  The lattice must be large enough to 
accomodate not only the polaron but also the boundary
layer.\cite{boundary}

\begin{figure}[ht]
\epsfxsize=2.5 in\centerline{\epsffile{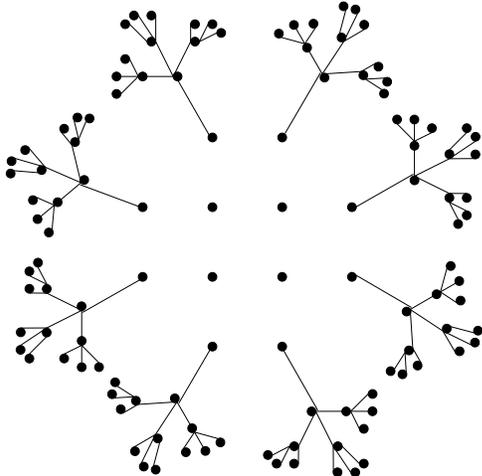}}
\caption{
A schematic diagram of a polaron with Bethe lattice ``hair".
}
\end{figure}

We also present some results on the $t-J^z$ model.  One might again 
expect the polaron picture to hold at small enough $J^z$.  However, 
we no longer have such a direct measurement of the polaron 
size as was afforded in the Heisenberg case by the total spin of
the groundstate, since $\vec S$ is not conserved in the Ising case.
We can still measure the total $z$-component, $S^z$.  However, it 
is reasonable to expect that the ferromagnetically aligned spins 
inside the polaron are oriented in the $xy$ plane, rather than along
the $z$-axis.  Classically this costs an exchange energy of $J/4$ 
per bond, rather than $J/2$ which would occur for $z$-orientation \cite{Barnes}.
Eqs. (\ref{E}), (\ref{E,R}) are thus modified to:
\begin{eqnarray}
E_M&=& (1/2)J_z\pi R^2\nonumber \\
R&=&-1.39 J_z^{-1/4}\nonumber \\
E&=& -4+6.03 \sqrt{J_z}.\end{eqnarray}
Our numerical results are consistent with this picture; we find
$S^z=0$ for all $J$.  A comparison of the polaron and BRSS energies 
to our DMRG results in shown in Fig. 6. We also show the  
formula of Barnes, et al.,
\begin{equation}
E\approx -3.63 + 2.93 J_z^{2/3},\end{equation}
obtained from fitting Lanczos and Monte Carlo data. 
 Perhaps surprisingly,
 the agreement with 
BRSS at larger $J$ is a bit worse than in the $t-J$ model.  
We also present $\langle S^z_{\vec r}S^z_{\vec r+\hat e}-n_{\vec r}
n_{\vec r+\hat e}/4\rangle$ in Fig. 7.  A central region where this is close 
to $-1/4$ is visible, corresponding to a polaron with spins lying
in the $xy$ plane. The slightly worse agreement with the polaron picture
in the $t-J^z$ model may just reflect the fact that the boundary
layer is wider leading to larger finite size corrections.   
A recent extensive discussion of the $t-J_z$ model may be found 
in Ref. \cite{Chernyshev}.
\begin{figure}[ht]
\epsfxsize=3.0 in\centerline{\epsffile{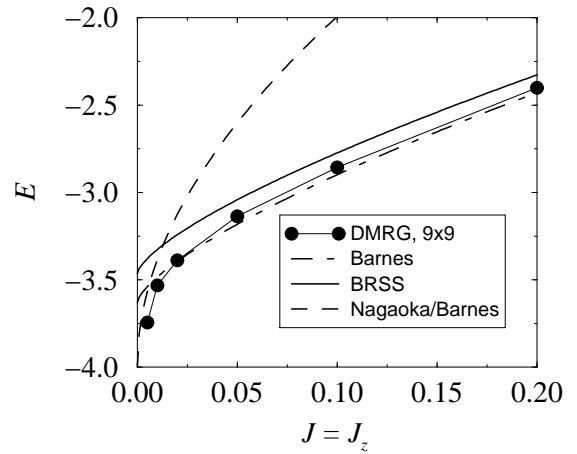}}
\caption{
For the $t-J_z$ model, the total energy of a hole, measured
relative to the undoped system, as a function of $J$, on a
$9\times9$ system. For comparison, we also show the approximate
energies
for a Nagaoka ferromagnetic polaron and for a hole according
to the BRSS treatment. The same extra term $J$ discussed in
the caption of Fig. 4 was added to each of the three analytic
curves.
}
\end{figure}         

\begin{figure}[ht]
\epsfxsize=2.5 in\centerline{\epsffile{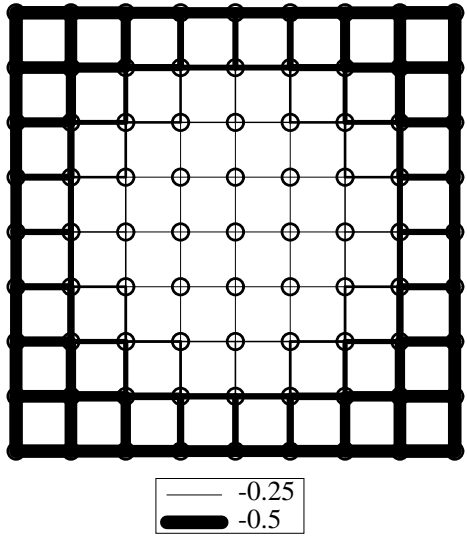}}
\caption{
For the $t-J_z$ model,
the expectation value of $ S^z_i S^z_j-n_in_j/4$ is
displayed as the thickness of each line, for a $9\times9$ system
with one hole and $J=0.01$. The ferromagnetic polaron, with
its magnetization direction in the $x-y$ plane, is clearly
visible in the central region, where the expectation value is
nearly -0.25.
}
\end{figure}         

We remark that the spin-orientation of the polaron is also a physical 
quantity in the $t-J$ model since the N\'eel order picks out a 
particular spin orientation ($\hat z$).  We might again expect that the 
polaron spin be oriented in the $xy$-plane, perpendicular to the
N\'eel order vector.  We may study this issue on a finite lattice
by applying a small staggered field at the boundary of our lattice.
The result, shown in Figs. 8 and 9, shows that indeed the polaron spin 
is oriented in the $xy$-plane. In Fig. 8, one sees that the
energy as a function of the total $z$ component of spin is
a minimum for $S_z=0$, corresponding to an orientation of the
polaron in the $x-y$ plane. The background antiferromagnetic
order is in the $z$ direction. The energy cost of
rotating the polaron towards the $z$ direction is small,
however. Once we take $S_z > 8$, the energy rises rapidly,
indicating that now the polaron is forced to change size, rather
than simply reorienting.
In Fig. 9, we show separately the
exchange and kinetic energies. The interpretation just discussed
based on the total energy in Fig. 8 is supported strongly by the sharp
kinks in the energies at $S_z = 8$ in Fig. 9. For $S_z > 8$, the
polaron is made larger, with a gain in $E_K$ but with a greater
cost in $E_M$.
One sees that for $S_z < 8$, the driving force
in orienting the polaron is the kinetic energy, which is lowest
for $S_z=0$. It appears that for an $x-y$ polaron, the hole is able to
penetrate more readily outside the ferromagnetic region,
lowering the kinetic energy.

\begin{figure}[ht]
\epsfxsize=2.9 in\centerline{\epsffile{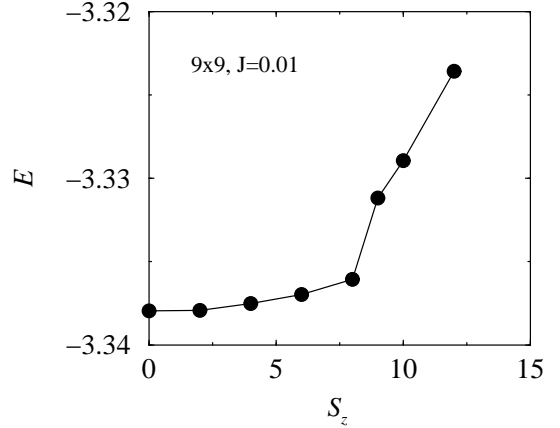}}
\caption{
The total energy of a hole, measured relative to the undoped
system,
as a function of $J$, on an open $9\times9$ system, with a
staggered
magnetic field of $\pm 0.1$ applied to all edge sites.
}
\end{figure}         

\begin{figure}[ht]
\epsfxsize=2.9 in\centerline{\epsffile{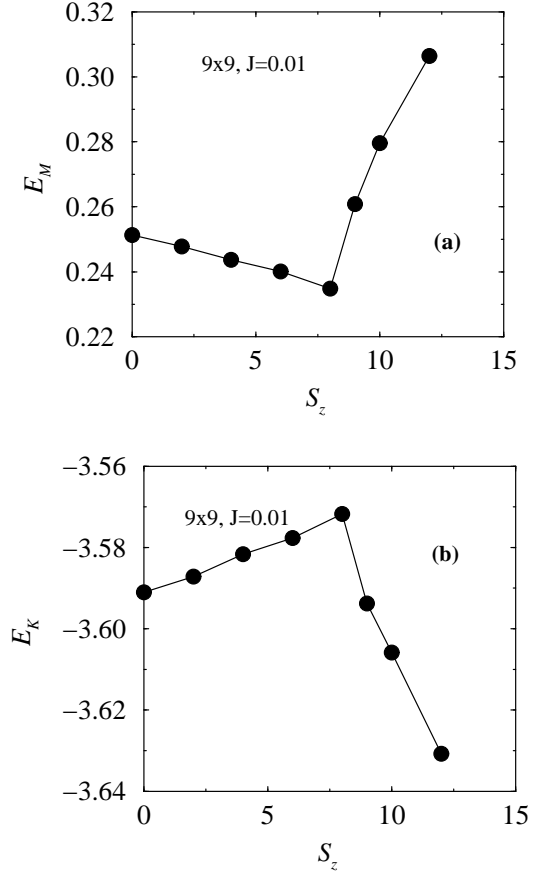}}
\caption{
(a) For the same system as in Fig. 8, the exchange energy of a
hole
as a function of $S_z$. (b) The kinetic energy of the hole. 
}
\end{figure}         

We thus conclude that the simple Nagaoka spin polaron picture for a single
hole in the $t-J$ model is valid for very small $J$, less than $0.02-0.03$,
and also is valid for the $t-J_z$ model. However, analytic estimates for
size of the polaron are not very accurate in the numerically accessible regime.
Furthermore, analytic estimates for the maximum $J$ at which it
a polaron exists are off by about a factor of five.
For larger $J$ the Brinkman-Rice picture of self-retracing
excursions into the surrounding antiferromagnetic region
gives energies in fairly good agreement with our results.

\acknowledgements This research is supported by the NSF under grant No.
DMR98-70930 (S.R.W.) and the NSF under grant No. PHY94-07195 and  NSERC of
 Canada (I.A.).  I.A. was a member of the Institute
 for Theoretical
Physics, University of Calfornia at Santa Barbara, during the time that
much of this research was carried out.  We would like to thank D.J. Scalapino
for helpful conversations.


\begin{references}
\bibitem{nagaoka} Y. Nagaoka, Phys. Rev. {\bf 147}, 392 (1966).

\bibitem{holestructures} S.R.~White and D.J.~Scalapino,
\prb  {\bf 55}, 6504 (1997).

\bibitem{dmrg} S.R. White, \prl {\bf 69}, 2863 (1992),
\prb {\bf 48}, 10345 (1993).

\bibitem{BR} W.F. Brinkman and T.M. Rice, Phys. Rev. {\bf B2}, 1324 (1970).

\bibitem{SS} B.I. Shraiman and E.D. Siggia, \prl {\bf 60}, 740
(1988).

\bibitem{Barnes} T. Barnes, E. Dagotto, A. Moreo and E.S. Swanson, 
Phys. Rev. {\bf B40}, 10977 (1989). See also J. Riera and 
E. Dagotto, Phys. Rev. {\bf B47}, 15346 (1993).

\bibitem{boundary}
If one wanted to improve the accuracy 
beyond the naive estimates associated with Eq. (5), one 
might try using a larger effective
radius for the kinetic energy than for exchange energy: 
$R_K = R + \Delta$, with $\Delta \approx 2$ representing the
thickness of the Bethe hair boundary layer. Then Eq. (5) becomes
\begin{equation}
E=-4+ (2.40)^2/(R+\Delta)^2+1.16J\pi R^2.
\end{equation}
Indeed, this does improve the agreement with the DMRG data (not
shown).

\bibitem{Chernyshev} A.L. Chernyshev and P.W. Leung, Phys. Rev. 
{\bf B60}, 1592 (1999).

\end{references}
\end{document}